\documentclass{kapproc} 
\usepackage{procps} 
\usepackage[dvips]{graphicx}

\upperandlowercase

\nochapequationnumber 

\let\footnote\savefootnote

\let\footnotetext\savefootnotetext

\normallatexbib 

\begin{document}
\articletitle{Neutral Dense Quark Matter}

\author{\underline{Mei Huang}}
\affil{Institut f\"ur Theoretische Physik, 
        J.W. Goethe-Universitaet, Germany}
\email{huang@th.physik.uni-frankfurt.de}

\author{Igor Shovkovy}
\affil{Institut f\"ur Theoretische Physik, 
        J.W. Goethe-Universitaet, Germany}
\email{shovkovy@th.physik.uni-frankfurt.de}

\begin{abstract} 
The ground state of dense up and down quark matter under local and global 
charge neutrality conditions with $\beta$-equilibrium has at least 
four possibilities: normal, regular 2SC, gapless 2SC phases, and mixed 
phase composed of 2SC phase and normal components. The discussion is focused 
on the unusual properties of gapless 2SC phase at zero as well as at finite 
temperature.
\end{abstract}


\section{Introduction}
Currently, our knowledge of sufficiently cold and dense matter is 
very limited. There are neither experimental data nor lattice data 
in this region. From the BCS theorem, it is speculated that if the matter is 
dense enough, the ground state of the deconfined quark matter at low 
temperature will be a color superconductor \cite{cs}. Recent studies
show that dense quark matter has a rich phase structure, 
see Ref. \cite{review} for reviews. 
At asymptotically high bayron densites, this phenomenon can be studied from 
first principles \cite{weak}. If all three colors of the three light quarks 
participate in the Cooper pairing, the ground state will be in the 
color-flavor-locking (CFL) phase \cite{cfl}.

In reality, we are more interested in the intermediate density region, where the 
color superconducting phase may exist in the interior of neutron stars 
or may be created in heavy ion collisions. Unfortunately, we have little knowledge
about this region: we are not sure how the deconfinement and the chiral restoration 
phase transitions happen, how the QCD coupling constant evolves and how the strange 
quark behaves in dense matter, etc. Primarily, our current understanding of the QCD 
phase structure in this region is based on assumptions. 

In the framework of the bag model \cite{bag-model}, or, in general, under the
assumption that the strange quark mass is small \cite{absence2sc}, one can exclude 
the 2SC phase in the interior of compact stars when charge neutrality is considered.

However, there is another possibility, if the strange quark becomes light 
at a larger chemical potential than the $u,d$ quarks, there will be a density region 
where only $u, d$ quarks exist. Here, we focus on the dense quark matter composed of 
only $u$ and $d$ quarks, assuming that the strange quark is too heavy to involve in 
the system. If bulk $u, d$ quark matter exists in the interior of compact stars, it 
should be neutral with respect to electrical as well as color charges 
\cite{absence2sc, huang-2sc, Blaschke-2sc, SH, misra, Ruster, neutral-buballa, SHH}. 
Also, such matter should remain in $\beta$-equilibrium, i.e., the chemical potential for 
each flavor and color should satisfy the relationship 
$\mu_{ij, \alpha\beta} = (\mu \delta_{ij}- \mu_e Q_{ij})\delta_{\alpha\beta} + 2/\sqrt{3} \mu_{8} \delta_{ij} (T_{8})_{\alpha \beta}$ , where $Q$ and $T_8$ are generators of U(1)$_{em}$ 
of electromagnetism and the U(1)$_{8}$ subgroup of the color gauge group. Satisfying these 
requirements imposes nontrivial relations between the chemical potentials of different 
quarks. We will see that these requiements play very important role in determining the ground 
state of dense $u,d$ quark matter. 

The charge neutrality condition can be satisfied locally 
\cite{absence2sc, huang-2sc,Blaschke-2sc,SH,misra,Ruster} or globally 
\cite{neutral-buballa, SHH}. In the following, we will firstly discuss the homogeneous phase 
when the charge neutrality is satisfied locally, then discuss the mixed phase when the charge 
neutrality condition is satisfied globally.

\section{Local charge neutrality: homogeneous phase}

\subsection{Correct way to find the neutral ground state}

To neutralize the electical charge in the homogeneous dense $u, d$ quark matter, 
roughly speaking, twice as many $d$ quarks as $u$ quarks are needed, i.e., $n_d \simeq 2 n_u$, 
where $n_{u,d}$ are the number densities for $u$ and $d$ quarks. This induces a mismatch
between the Fermi surfaces of pairing quarks, i.e., $\mu_d - \mu_u = \mu_e = 2 \delta\mu$, where
$\mu_{e}$ is the electron chemical potential. 

To get the ground state of the system, we need to know the thermodynamical potential. 
For simplicity, we use Nambu--Jona-Lasinio (NJL) model \cite{SKP} to describe 2-flavor 
quark matter, 
\begin{eqnarray}
\label{lagr}
{\cal L} & = &\bar{q} i\gamma^{\mu}\partial_{\mu} q + 
 G_S\left[(\bar{q}q)^2 + (\bar{q}i\gamma_5{\bf \vec{\tau}}q)^2\right] 
\nonumber \\
 &&+G_D\left[(i \bar{q}^C  \varepsilon  \epsilon^{b} \gamma_5 q )
 (i \bar{q} \varepsilon \epsilon^{b} \gamma_5 q^C)\right],
\end{eqnarray}
where $q^C=C {\bar q}^T$ is the charge-conjugate spinor and $C=i\gamma^2
\gamma^0$ is the charge conjugation matrix. The quark field $q \equiv
q_{i\alpha}$ is a four-component Dirac spinor that carries flavor
($i=1,2$) and color ($\alpha=1,2,3$) indices. The Pauli matrices
are denoted by ${\vec \tau} =(\tau^1,\tau^2, \tau^3)$, while
$(\varepsilon)^{ik} \equiv \varepsilon^{ik}$ and $(\epsilon^b)^{\alpha
\beta} \equiv \epsilon^{\alpha \beta b}$ are the antisymmetric tensors 
in the flavor and color spaces, respectively. We also introduce a 
momentum cutoff $\Lambda$, and two independent coupling constants in 
the scalar quark-antiquark and scalar diquark channels, $G_S$ and $G_D$.
We define $\eta=G_D/G_S$, and $\eta=0.75$ is the value from the Fierz 
transformation, but in principle, $\eta$ is a free parameter.

In the mean-field approximation, the thermodynamical potential for $u, d$ quarks in 
$\beta$-equilibrium with electrons takes the form \cite{huang-2sc, SH}:
\begin{eqnarray} 
\Omega_{u,d,e} &=& \Omega_{0}
-\frac{1}{12\pi^2}\left(\mu_{e}^{4}+2\pi^{2}T^{2}\mu_{e}^{2}
+\frac{7\pi^{4}}{15} T^{4} \right) + \frac{m^2}{4G_S} \nonumber\\
&+&  \frac{\Delta^2}{4G_D}
-\sum_{a} \int\frac{d^3 p}{(2\pi)^3} \left[E_{a}
+2 T\ln\left(1+e^{-E_{a}/T}\right)\right] ,
\label{pot}
\end{eqnarray} 
where $\Omega_{0}$ is a constant added to make the pressure of the vacuum 
zero, and the electron mass was taken to be zero, which is sufficient
for the purposes of the current study. The sum in the second line of
Eq.~(\ref{pot}) runs over all (6 quark and 6 antiquark)
quasi-particles. The explicit dispersion relations and the degeneracy 
factors of the quasi-particles read
\begin{eqnarray}
E_{ub}^{\pm} &=& E(p) \pm \mu_{ub} , \hspace{26.6mm} [\times 1]
\label{disp-ub} \\
E_{db}^{\pm} &=& E(p) \pm \mu_{db} , \hspace{26.8mm} [\times 1]
\label{disp-db}\\
E_{\Delta^{\pm}}^{\pm} &=& E_{\Delta}^{\pm}(p) \pm  \delta \mu .
\hspace{25.5mm} [\times 2]
\label{2-degenerate}
\end{eqnarray}
Here we introduced the following shorthand notation:
$E(p) = \sqrt{{\bf p}^2+m^2}$ and $ E_{\Delta}^{\pm}(p) = 
\sqrt{[E(p) \pm \bar{\mu}]^2 +\Delta^2}$
with $\bar{\mu} \equiv \mu- \mu_{e}/6+ \mu_{8}/3$. 

If a macroscopic chunk of quark matter is created in heavy ion collisions
or exists inside the compact stars, it must be in color singlet. So in the 
following discussions, color charge neutrality condition is always
satisfied. 

Now, we discuss the role of electrical charge neutrality condition. 
If a macroscopic chunk of quark matter has nonzero net electrical charge density $n_Q$,
the total thermodynamical potential for the system should be given by
\begin{eqnarray}
\Omega &=&  \Omega_{Coulomb} + \Omega_{u,d,e}, 
\end{eqnarray}
where $\Omega_{Coulomb} \sim n_Q^2 V^{2/3}$ ($V$ is the volume of the 
system) is induced by the repulsive Coulomb interaction. The energy 
density grows with increasing the volume of the system, as a result, it is almost
impossible for matter inside stars to remain charged over macroscopic distances.
So the bulk quark matter should also satisfy electrical neutrality condition,
thus $\Omega_{Coulomb}|_{n_Q=0}=0$, and $\Omega_{u,d,e}|_{n_Q=0}$ 
is on the neutrality line. Under the charge neutrality condition, the total thermodynamical 
potential of the system is $\Omega|_{n_Q=0}=\Omega_{u,d,e}|_{n_Q=0}$. 
 
Here, we want to emphasize that: {\it  The correct way to find the ground state of 
the homogeneous neutral $u, d$ quark matter is to minimize the thermodynamical 
potential along the neutrality line $\Omega|_{n_Q=0} = \Omega_{u,d,e}|_{n_Q=0}$, not 
like in the flavor asymmetric quark system, where $\beta$-equilibrium is required but 
$\mu_e$ is a free parameter, and the ground state is determined 
by minimizing the thermodynamical potential $\Omega_{u,d,e}$.}

\begin{figure}
\includegraphics[bbllx=110,bblly=8,bburx=480,bbury=250,width=8cm]
{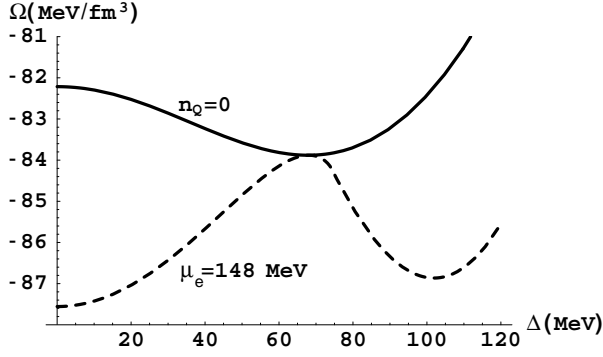}
\caption{The effective potential as a function of the diquark gap $\Delta$
calculated at a fixed value of the electrical chemical potential $\mu_e =
148 $ ~MeV (dashed line), and the effective potential defined along the
neutrality line (solid line). The results are plotted for $\mu=400$ MeV
with $\eta=0.75$.}
\label{V2D}
\end{figure}

From Figure 1, we can see the difference in determining the ground state for a charge 
neutral system and for a flavor asymmetric system. In Figure 1, at a given chemical 
potential $\mu=400 ~ {\rm MeV}$ and $\eta=0.75$, the thermodynamical potential 
along the charge neutrality line $\Omega|_{n_Q=0}$ as a function of the diquark 
gap $\Delta$ is shown by the solid line. The minimum gives the ground state of the neutral 
system, and the corresponding values of the chemical potential and the diquark gap are 
$\mu_e=148 ~{\rm MeV}$ 
and $\Delta=68 ~{\rm MeV}$, respectively. If we switch off the charge neutrality conditions, and 
consider the flavor asymmetric $u, d$ quark matter in $\beta-$ equilibrium \cite{asymmetric}, 
the electrical chemical potential $\mu_e$ becomes a free parameter. 
At a fixed $\mu_e=148 ~{\rm MeV}$ and with color charge neutrality, the thermodynamical 
potential is shown as a function of the diquark gap by the dashed line in Figure 1. 
The minimum gives the ground state of the flavor asymmetric system, and the corresponding 
diquark gap is $\Delta=104 ~{\rm MeV}$, but this state has positive electrical charge density,
and cannot exist in the interior of compact stars.

\subsection{$\eta$ dependent solutions}

In the last subsection, by looking for the minimum of the thermodynamic potential along the 
charge neutrality line, we found the ground state for the charge neutral $u, d$ quark system. 

Equivalently, the neutral ground state can also be determined by solving the diquark gap 
equation together with the charge neutrality conditions. We visualize this method in Figure 2, with
color neutrality always satisfied, at a given chemical potential $\mu=400 ~{\rm MeV}$.
The nontrivial solutions to the diquark gap equation as functions of the electrical chemical 
potential $\mu_e$ are shown by a thick-solid line ($\eta=0.75$), a long-dashed line ($\eta=1.0$),
and a short-dashed line ($\eta=0.5$). It is found that for each $\eta$, the solution is divided 
into two branches by the thin-solid line $\Delta=\delta\mu$, and 
the solution is very sensitive to $\eta$. Also, there is always a trivial solution to the diquark gap 
equation, i.e., $\Delta =0$.  The solution of the charge neutrality conditions is shown by a 
thick dash-dotted line, which is also divided into two branches by the thin-solid line 
$\Delta=\delta\mu$, but the solution of the charge neutrality is independent of $\eta$.

\begin{figure}
\includegraphics[bbllx=88,bblly=4,bburx=572,bbury=308,width=8cm]
{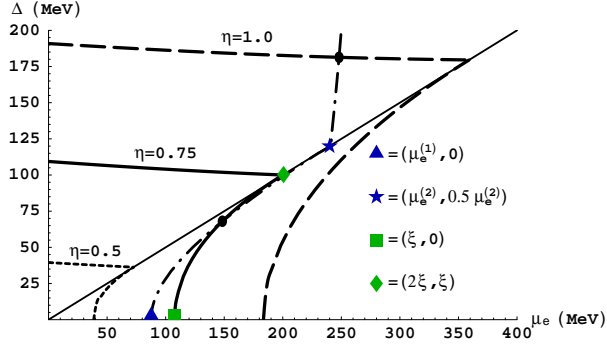}
\caption{The graphical representation of the solution to the charge 
neutrality conditions (thick dash-dotted line) and the solution to 
the gap equation for three different values of the diquark coupling 
constant (thick solid and dashed lines). The intersection points 
represent the solutions to both. The thin solid line divides two 
qualitatively different regions, $\Delta<\delta\mu$ and $\Delta>
\delta\mu$. The results are plotted for $\mu=400$ MeV and three 
values of diquark coupling constant $G_{D} = \eta G_S$ with $\eta=0.5$, 
$\eta=0.75$, and $\eta=1.0$.}
\label{gapneutral}
\end{figure}

The cross-point of the solutions to the charge neutrality conditions and the diquark gap 
gives the solution of the system. We find that the neutral ground state is sensitive to 
the coupling constant $G_D=\eta G_S$ in the diquark channel. In the case of a very strong coupling 
(e.g., $\eta=1.0$ case), the charge neutrality line crosses the upper branch of the solution to the 
diquark gap, the ground state is a charge neutral regular 2SC phase with $\Delta>\delta\mu$. 
In the case of weak coupling (e.g., $\eta=0.5$), the charge neutrality line crosses the trivial 
solution of the diquark gap, i.e., the ground state is a charge neutral normal quark matter with 
$\Delta=0$. The regime of intermediate coupling (see, e.g., $\eta=0.75$ case) is most interesting,
the charge neutrality line crosses the lower branch of the solution of the diquark gap. We will 
see that this phase is a gapless 2SC (g2SC) phase with $\Delta<\delta\mu$, which is different from 
the regular 2SC phase, and has some unusual properties.   

\subsection{g2SC phase}

In this subsection, we will explain why we call the color superconducting phase 
with $\Delta<\delta\mu$ the g2SC phase, and we will show some special properties of
this phase. 

\vspace{3mm}
{\bf Quasi-particle spectrum} 
\vspace{3mm}

It is instructive to start with the excitation spectrum in the case of the ordinary 2SC phase 
when $\delta\mu=0$. With the conventional choice of the gap pointing in the anti-blue direction 
in color space, the blue quarks are not affected by the pairing dynamics, and the other four 
quarsi-particle excitations are linear superpositions of $u_{r,g}$ and $d_{r,g}$
quarks and holes. The quasi-particle is nearly identical with a quark at large momenta
and with a hole at small momenta. We represent the quasi-particle in the form 
of $Q(quark, hole)$, then the four quasi-particles can be represented explicitly as 
$Q(u_r, d_g)$, $Q(u_g, d_r)$, $Q(d_r, u_g)$ and $Q(d_g, u_r)$. When $\delta\mu=0$, the four 
quasi-particles are degenerate, and have a common gap $\Delta$.

\begin{figure}
\hbox{
\includegraphics[bbllx=88,bblly=4,bburx=561,bbury=504,width=6cm]
{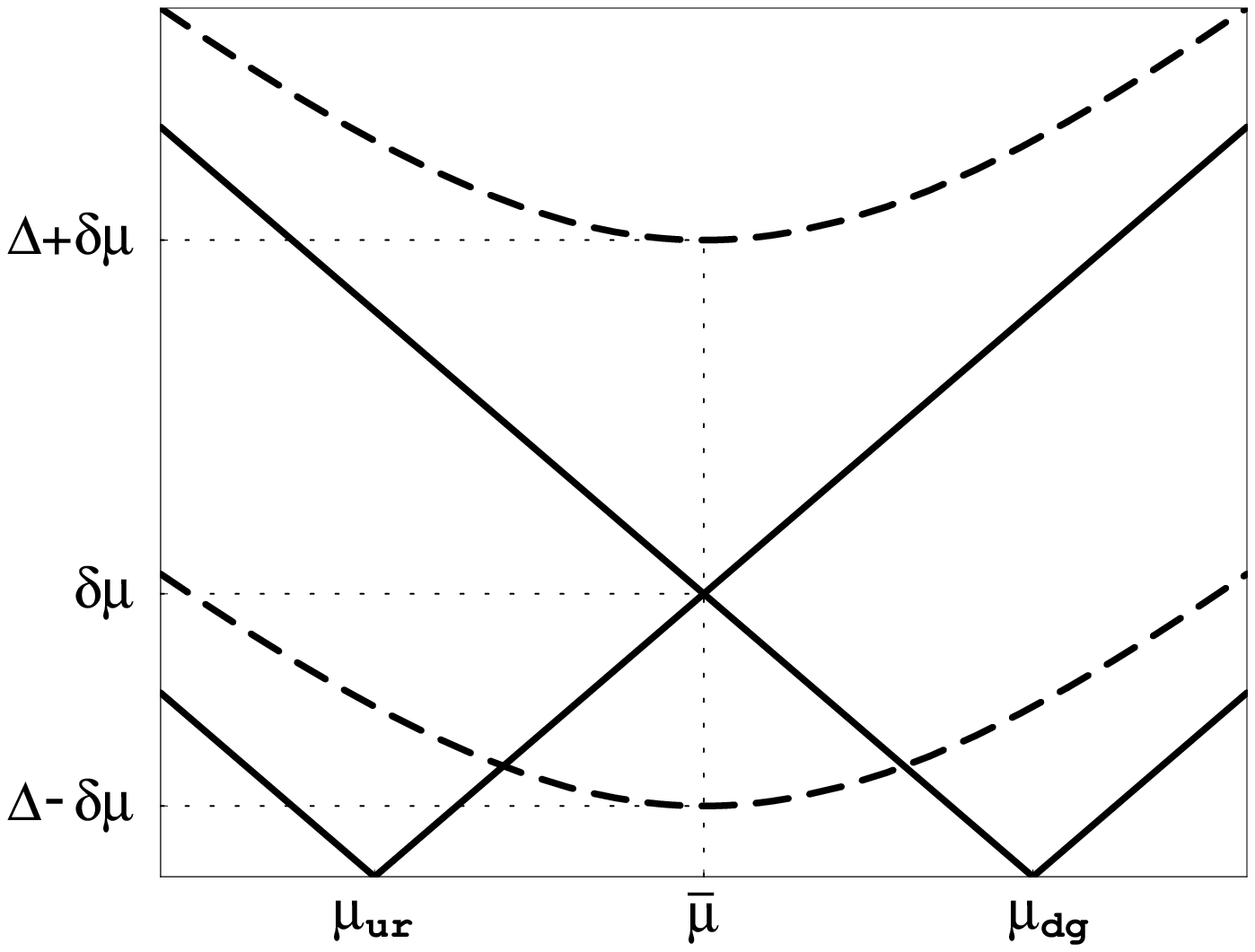}
\includegraphics[bbllx=88,bblly=4,bburx=561,bbury=504,width=6cm]
{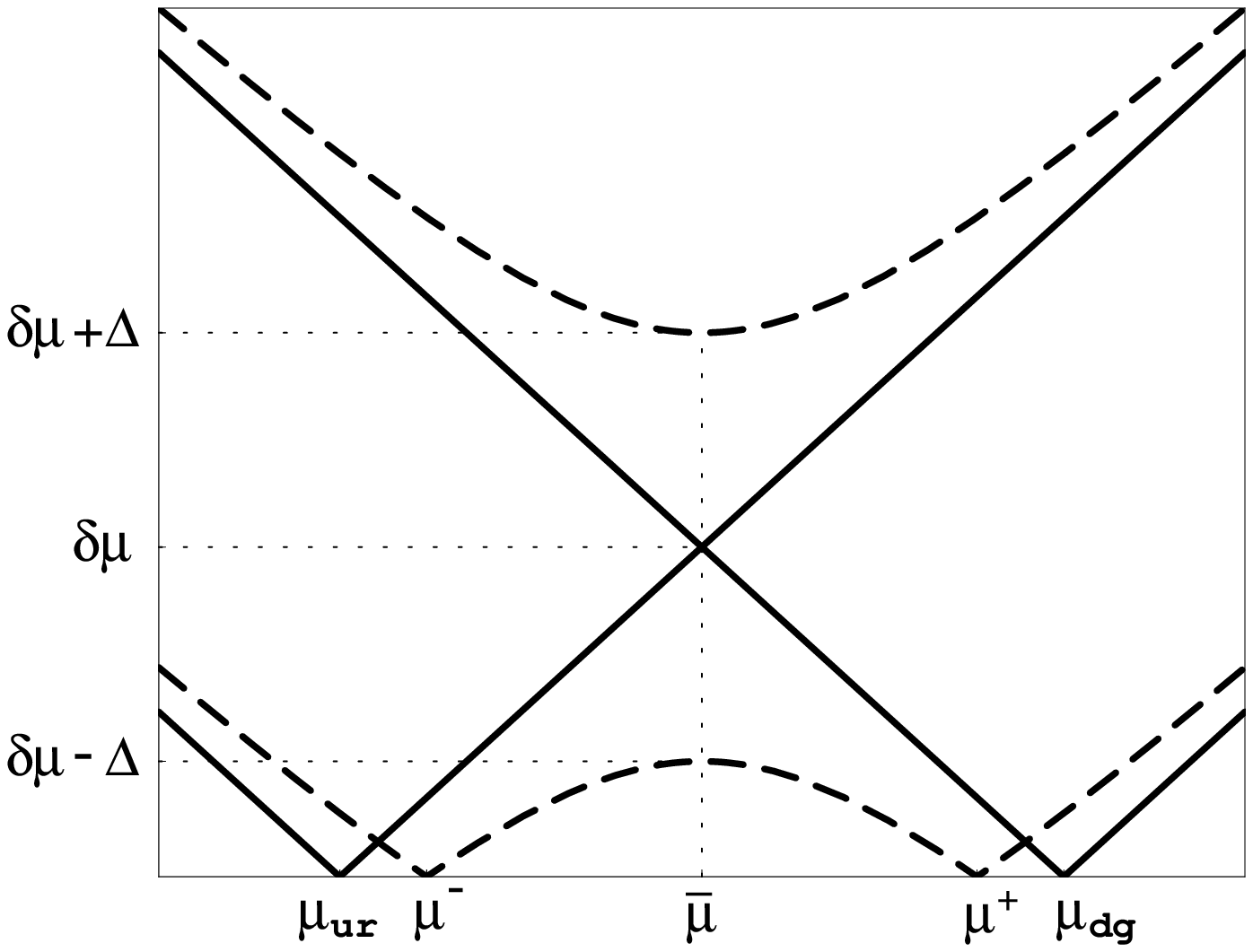}
}
\caption{The quasi-particle dispersion relations at low energies in the 2SC phase 
(left panel) and in the g2SC phase (right panel).  }
\end{figure}


If there is a small mismatch ($\delta\mu < \Delta$) between the Fermi surfaces of the pairing
$u$ and $d$ quarks, the excitation spectrum will change. For example, we show the excitation
spectrum of $Q(u_r, d_g)$ and $Q(d_g, u_r)$ in the left panel of  Figure 3. We can see that 
$\delta \mu$ induces two different dispersion relations, the quasi-particle $Q(d_g, u_r)$ has a 
smaller gap $\Delta - \mu$, and the quasi-particle $Q(u_r, d_g)$ has a larger gap $\Delta + \mu$.
This is similar to the case when the mismatch is induced by the mass difference of the pairing 
quarks \cite{gaplessCFL}.

If the mismatch $\delta\mu$ is larger than the gap parameter $\Delta$, the lower dispersion relation
for the quasi-particle $Q(d_g, u_r)$ will cross the zero-energy axis, as shown in the right panel 
of Figure 3. The energy of the quasi-particle $Q(d_g, u_r)$ vanishes at two values of momenta 
$p=\mu^{-}$ and $p=\mu^{+}$ where $\mu^{\pm}\equiv \bar\mu\pm\sqrt{(\delta\mu)^2-\Delta^2}$. 
This is why we call this phase gapless 2SC (g2SC) phase. An unstable gapless CFL phase has been  
found in Ref. \cite{gaplessCFL}, and a similar stable gapless color superconductivity
could also appear in a cold atomic gas or in $u, s$ or $d, s$ quark matter when the  
number densities are kept fixed \cite{GLW}.

As one would expect, far outside the pairing region, $p\simeq\bar\mu$, 
the quasi-particle dispersion relations are similar to those in the 2SC 
phase. Also, around $p\simeq\bar\mu$, the quasi-particle $Q(u_r,d_g)$ resembles the 
dispersion relations with that in the regular 2SC phase. The most remarkable property of the 
quasi-particle spectra in the g2SC phase is that the low energy excitations 
($E\ll \delta\mu-\Delta$) are very similar to those in the normal phase represented 
by solid lines. The only difference is that the values of the chemical potentials of 
the up and down quarks $\mu_{ur}=\mu_{ug}$ and $\mu_{dg}=\mu_{dr}$ are replaced 
by the values $\mu^{-}$ and $\mu^{+}$, respectively. This observation 
suggests, in particular, that the low energy (large distance scale) 
properties of the g2SC phase should look similar to those in the 
normal phase.   

\vspace{3mm}
{\bf Finite temperature properties}
\vspace{3mm}

In a superconducting system, when one increases the temperature at a given 
chemical potential, thermal motion will eventually break up the quark 
Cooper pairs. In the weakly interacting Bardeen-Copper-Schrieffer (BCS) 
theory, the transition between the superconducting and normal phases is 
usually of second order. The ratio of the critical temperature 
$T_c^{\rm BCS}$ to the zero temperature value of the gap 
$\Delta_0^{\rm BCS}$ is a universal value \cite{ratio-in-BCS}
\begin{eqnarray}
r_{\rm BCS}=\frac{T_c^{\rm BCS}}{\Delta_0^{\rm BCS}} =
\frac{{\rm e}^{\gamma_E}}{\pi} \approx 0.567,
\label{r_BCS}
\end{eqnarray}
where $\gamma_E \approx 0.577$ is the Euler constant.
In the conventional 2SC phase of quark matter with equal densities of
the up and down quarks, the ratio of the critical temperature to the 
zero temperature value of the gap is also the same as in the BCS theory 
\cite{PR-sp1}. In the spin-0 color flavor locked phase as well as 
in the spin-1 color spin locked phase, on the other hand, this ratio
is larger than the BCS ratio by the factors $2^{1/3}$ and $2^{2/3}$, 
respectively. These deviations are related directly to the presence 
of two different types of quasi-particles with nonequal gaps 
\cite{ratio-in-CFL}. 

\begin{figure}
\includegraphics[bbllx=100,bblly=12,bburx=571,bbury=310,width=8cm]
{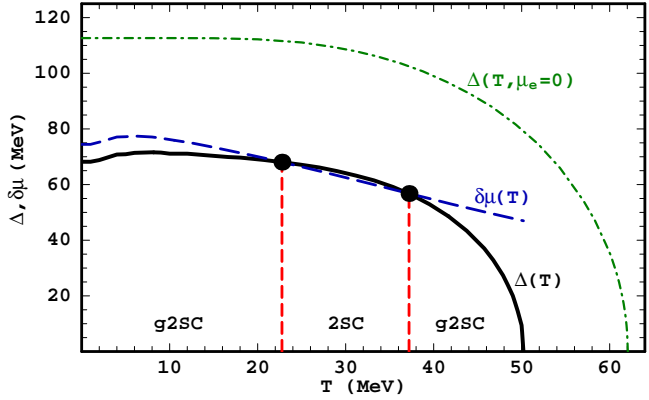}
\caption{The temperature dependence of the diquark gap (solid line) 
and the value of $\delta\mu\equiv \mu_e/2$ (dashed line) in neutral 
quark matter. For comparison, the diquark gap in the model with 
$\mu_e=0$ and $\mu_8=0$ is also shown (dash-dotted line). The results 
are plotted for $\mu=400$ MeV and $\eta=0.75$.}

\label{gap-mue0-vs-T.eps}
\end{figure}

For the g2SC phase, 
the typical results for the default choice of parameters $\mu=400$ MeV 
and $\eta=0.75$ are shown in Figure~\ref{gap-mue0-vs-T.eps}. Both the 
values of the diquark gap (solid line) and the mismatch parameter 
$\delta\mu=\mu_e/2$ (dashed line) are plotted. One very unusual property 
of the shown temperature dependence of the gap is a nonmonotonic behavior.
Only at sufficiently high temperatures, the gap is a decreasing function. 
In the low temperature region, $T\leq 10$ MeV, however, it increases 
with temperature. For comparison, in the same figure, the diquark gap 
in the model with $\mu_e=0$ and $\mu_8=0$ is also shown (dash-dotted 
line). This latter has the standard BCS shape.

Another interesting thing regarding the temperature dependences in
Figure~\ref{gap-mue0-vs-T.eps} appears in the intermediate temperature 
region, $22.5 \leq T \leq 37$ MeV. By comparing the values of 
$\Delta(T)$ and $\delta\mu$ in this region, we see that the g2SC phase 
is replaced by a ``transitional'' 2SC phase there. Indeed, the 
energy spectrum of the
quasi-particles even at finite temperature is determined by the same 
relations in Eqs.~(\ref{disp-ub}) and (\ref{2-degenerate}) that we 
used at zero temperature. When $\Delta > \delta \mu$, the modes 
determined by Eq.~(\ref{2-degenerate}) are gapped. Then, according 
to our standard classification, the ground state is the 2SC phase.

It is fair to say, of course, that the qualitative difference of the 
g2SC and 2SC phases is not so striking at finite temperature as it 
is at zero temperature. This difference is particularly negligible 
in the region of interest where temperatures $22.5 \leq T 
\leq 37$ MeV are considerably larger than the actual value of 
the smaller gap, $\Delta - \delta \mu$. However, by increasing the value 
of the coupling constant slightly, the transitional 2SC phase 
can be made much stronger and the window of intermediate temperatures 
can become considerably wider. In either case, we find it 
rather unusual that the g2SC phase of neutral quark matter is 
replaced by a transitional 2SC phase at intermediate temperatures 
which, is replaced by the g2SC phase again at higher temperatures. 

\begin{figure}
\includegraphics[bbllx=110,bblly=14,bburx=590,bbury=315,width=8cm]
{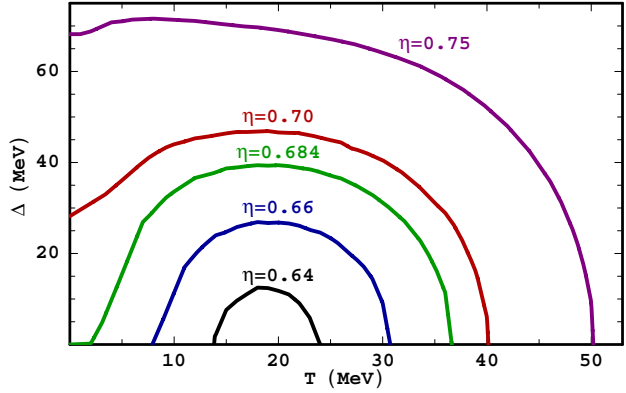}
\caption{The temperature dependence of the diquark gap in neutral 
quark matter calculated for several values of the diquark coupling 
strength $\eta=G_D/G_S$.}

\label{gap-eta}
\end{figure}

It appears that the temperature dependence of the diquark gap is 
very sensitive to the choice of the diquark coupling strength 
$\eta=G_D/G_S$ in the model at hand. This is not surprising because 
the solution to the gap equation is very sensitive to this choice. 
The resulting interplay of the solution for $\Delta$ with the 
condition of charge neutrality, however, is very interesting.
This is demonstrated by the plot of the temperature dependence 
of the diquark gap calculated for several values of the diquark 
coupling constant in Figure~\ref{gap-eta}.

The most amazing are the results for weak coupling. It appears that
the gap function could have sizable values at finite temperature even
if it is exactly zero at zero temperature. This possibility comes
about only because of the strong influence of the neutrality condition
on the ground state preference in quark matter. Because of the thermal
effects, the positive electrical charge of the diquark condensate is
easier to accommodate at finite temperature. We should mention that 
somewhat similar results for the temperature dependence of the gap were 
also obtained in Ref.~\cite{SedLom} in a study of the asymmetric nuclear 
matter, and in Ref. \cite{liaojf} when number density was fixed.

\begin{figure}
\includegraphics[bbllx=100,bblly=11,bburx=590,bbury=314,width=8cm]
{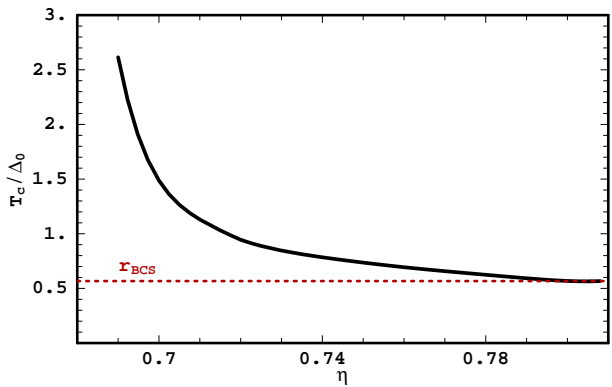}
\caption{The ratio of the critical temperature to the zero temperature 
gap in neutral quark matter as a function of the coupling strength
$\eta=G_D/G_S$.}
\label{ratio}
\end{figure}

The numerical results for the ratio of the critical temperature to 
the zero temperature gap in the g2SC case as a function of the diquark 
coupling strength $\eta=G_D/G_S$ are plotted in Figure~\ref{ratio}. 
The dependence is shown for the most interesting range of values of 
$\eta=G_D/G_S$, $0.68 \leq \eta \leq 0.81$, which allows 
the g2SC stable ground state at zero temperature. When the coupling 
gets weaker in this range,
the zero temperature gap vanishes gradually. As we saw from 
Figure~\ref{gap-eta}, however, this does not mean that the critical 
temperature vanishes too. Therefore, the ratio of a finite value of 
$T_c$ to the vanishing value of the gap can become arbitrarily 
large. In fact, it remains strictly infinite for a range of 
couplings.  

\section{Global charge neutrality: mixed phase}

We have discussed the homogeneous 2-flavor quark matter when charge neutrality
conditions are satisfied locally, and found that the local charge neutrality 
conditions impose very strong constraints on determining the ground state of the 
system.

On the other hand, one can construct mixed phase when charge neutrality conditions 
are satisfied globally. Inside mixed phases, the charge neutrality is satisfied 
``on average'' rather than locally. This means that different components of mixed 
phases may have non-zero densities of conserved charges, but the total charge of
all components still vanishes. In this case, one says that the local
charge neutrality condition is replaced by a global one. There are three possible 
components: (i) normal phase, (ii) 2SC phase, and (iii) g2SC phase. 

\begin{figure}
\includegraphics[bbllx=90,bblly=6,bburx=550,bbury=504,width=6cm]
{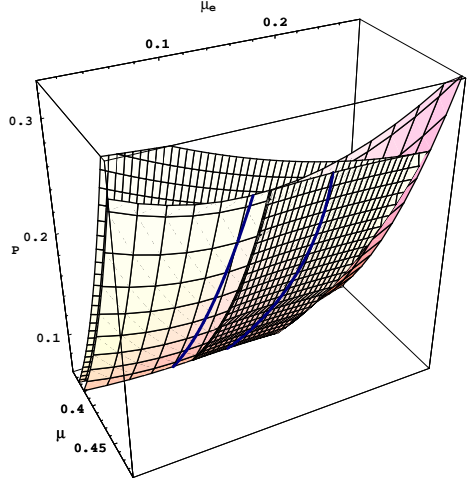}
\caption{\label{fig-back}
At $\eta=0.75$, pressure as a function of $\mu\equiv\mu_B/3$ and $\mu_e$ for 
the normal and color superconducting quark phases. The dark solid lines 
represent two locally neutral phases: (i) the neutral normal quark phase on the
left, and (ii) the neutral gapless 2SC phase on the right. The appearance
of the swallowtail structure is related to the first order type of the
phase transition in quark matter.}
\end{figure}

The pressure of the main three phases of two-flavor quark matter 
as a function of the baryon and electrical chemical potentials is shown in Figure
\ref{fig-back} at $\eta=0.75$. In this figure, we also show the pressure of the 
neutral normal quark and gapless 2SC phases (two dark solid lines). 
The surface of the g2SC phase extends only over a finite range of the values of 
$\mu_{e}$. It merges with the pressure surfaces of the normal quark phase 
(on the left) and with the ordinary 2SC phase (on the right).

It is interesting to notice that the three pressure surfaces in Figure  
\ref{fig-back} form a characteristic swallowtail structure. As one
could see, the appearance of this structure is directly related to the
fact that the phase transition between color superconducting and 
normal quark matter, which is driven by changing parameter $\mu_{e}$, is
of first order. In fact, one should expect the appearance of a similar
swallowtail structure also in a self-consistent description of the
hadron-quark phase transition. Such a description, however, is not
available yet.

From Figure \ref{fig-back}, one could see that the surfaces of normal  
and 2SC quark phases intersect along a common line. This means that the      
two phases have the same pressure along this line, and therefore could
potentially co-exist. Moreover, as is easy to check, normal quark
matter is negatively charged, while 2SC quark matter is positively
charged on this line.  This observation suggests that the appearance of   
the corresponding mixed phase is almost inevitable. 

Let us start by giving a brief introduction into the general method of
constructing mixed phases by imposing the Gibbs conditions of equilibrium
\cite{glen92,Weber}. From the physical point of view, the Gibbs
conditions enforce the mechanical as well as chemical equilibrium between
different components of a mixed phase. This is achieved by requiring that
the pressure of different components inside the mixed phase are equal,
and that the chemical potentials ($\mu$ and $\mu_{e}$) are the same
across the whole mixed phase. For example, in relation to the mixed phase
of normal and 2SC quark matter, these conditions read
\begin{eqnarray}
P^{(NQ)}(\mu,\mu_{e}) &=& P^{(2SC)}(\mu,\mu_{e}), 
\label{P=P}\\ 
\mu &=& \mu^{(NQ)}=\mu^{(2SC)}, 
\label{mu=mu}\\ 
\mu_{e} &=& \mu^{(NQ)}_{e}=\mu^{(2SC)}_{e}.  
\label{mue=mue}
\end{eqnarray}
 
\begin{figure}
\includegraphics[bbllx=88,bblly=4,bburx=561,bbury=504,width=6cm]
{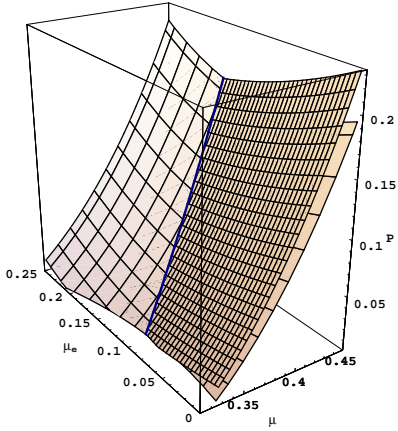}
\caption{\label{fig-front-quark}
At $\eta=0.75$, pressure as a function of $\mu\equiv\mu_B/3$ and $\mu_e$ for 
the normal and color superconducting quark phases (the same as in 
Figure \ref{fig-back}, but from a different viewpoint). The dark solid line
represents the mixed phase of negatively charged normal quark matter and
positively charged 2SC matter.}
\end{figure}

It is easy to visualize these conditions by plotting the pressure as a
function of chemical potentials ($\mu$ and $\mu_{e}$) for both components
of the mixed phase. This is shown in Figure \ref{fig-front-quark}. As
should be clear, the above Gibbs conditions are automatically satisfied
along the intersection line of two pressure surfaces (dark solid line in
Figure \ref{fig-front-quark}).

Different components of the mixed phase occupy different volumes of
space. To describe this quantitatively, we introduce the volume fraction
of normal quark matter as follows: $\chi^{NQ}_{2SC}\equiv V_{NQ}/V$
(notation $\chi^{A}_{B}$ means volume fraction of phase A in a mixture
with phase B). Then, the volume fraction of the 2SC phase is given by
$\chi^{2SC}_{NQ}=(1-\chi^{NQ}_{2SC})$. From the definition, it is clear
that $0\leq \chi^{NQ}_{2SC} \leq 1$.

The average electrical charge density of the mixed phase is determined by
the charge densities of its components taken in the proportion of the
corresponding volume fractions. Thus,
\begin{equation}
n^{(MP)}_{e} = \chi^{NQ}_{2SC} n^{(NQ)}_{e}(\mu,\mu_e) 
+(1-\chi^{NQ}_{2SC}) n^{(2SC)}_{e}(\mu,\mu_e).
\end{equation}
If the charge densities of the two components have opposite signs, one
can impose the global charge neutrality condition, $n^{(MP)}_{e}=0$.
Otherwise, a neutral mixed phase could not exist. In the case of 
quark matter, the charge density of the normal quark phase is negative,
while the charge density of the 2SC phase is positive along the line of
the Gibbs construction (dark solid line in Figure \ref{fig-front-quark}).
Therefore, a neutral mixed phase exists. The volume fractions of its
components are
\begin{eqnarray}
\chi^{NQ}_{2SC} &=& \frac{n^{(2SC)}_{e}}{n^{(2SC)}_{e}-n^{(NQ)}_{e}}, \\
\chi^{2SC}_{NQ} &\equiv& 1-\chi^{NQ}_{2SC}=
\frac{n^{(NQ)}_{e}}{n^{(NQ)}_{e}-n^{(2SC)}_{e}}.
\end{eqnarray}

After the volume fractions have been determined from the condition of the
global charge neutrality, we could also calculate the energy density of
the corresponding mixed phase,
\begin{equation}
\varepsilon^{(MP)} = \chi^{NQ}_{2SC} \varepsilon^{(NQ)}(\mu,\mu_e)
+(1-\chi^{NQ}_{2SC}) \varepsilon^{(2SC)}(\mu,\mu_e).
\end{equation}
This is essentially all that we need in order to construct the equation of
state of the mixed phase. 

So far, we were neglecting the effects of the Coulomb forces and the
surface tension between different components of the mixed phase. In a real
system, however, these are important. In particular, the balance
between the Coulomb forces and the surface tension determines the
size and geometry of different components inside the mixed phase. 

In our case, nearly equal volume fractions of the two quark phases are
likely to form alternating layers (slabs) of matter. The energy cost per
unit volume to produce such layers scales as $\sigma^{2/3}
(n_{e}^{(2SC)}-n_{e}^{(NQ)})^{2/3}$ where $\sigma$ is the surface tension
\cite{geometry}. Therefore, the quark mixed phase is a favorable phase of
matter only if the surface tension is not too large. Our simple estimates
show that $\sigma_{max} \leq 20$ MeV/fm$^{2}$. However, even for slightly
larger values, $20 \leq \sigma \leq 50$ MeV/fm$^{2}$, the mixed phase is
still possible, but its first appearance would occur at larger densities,
$3\rho_0 \leq \rho_B \leq 5\rho_0$. The value of the maximum surface
tension obtained here is comparable to the estimate in the case of the
hadronic-CFL mixed phase obtained in Ref. \cite{interface}. The thickness
of the layers scales as $\sigma^{1/3} (n_{e}^{(2SC)}-n_{e}^{(NQ)})^{-2/3}$
\cite{geometry}, and its typical value is of order $10$ fm in the quark
mixed phase. This is similar to the estimates in various hadron-quark and
hadron-hadron mixed phases \cite{geometry,interface}. While the actual
value of the surface tension in quark matter is not known, in this study
we assume that it is not very large. Otherwise, the homogeneous gapless
2SC phase should be the most favorable phase of nonstrange quark matter
\cite{SH}.

Under the assumptions that the effect of Coulomb forces and the surface tension is 
small, the mixed phase of normal and 2SC
quark matter is the most favorable neutral phase of matter in the model
at hand with $\eta=0.75$. This should be clear from observing the pressure surfaces in
Figs. \ref{fig-back} and \ref{fig-front-quark}. For a given value of the
baryon chemical potential $\mu=\mu_{B}/3$, the mixed phase is more
favorable than the gapless 2SC phase, while the gapless 2SC phase is more
favorable than the neutral normal quark phase.

\section{Conclusion}

Dense $u,d$ quark matter under local and global charge neutrality conditions 
in $\beta$-equilibrium has been discussed. 

Under local charge neutrality condition, the homogeneous neutral ground state 
is sensitive to the coupling constant in the diquark channel, it will be 
in the regular 2SC phase when the coupling is strong, in the normal phase
when the coupling is weak, and in the g2SC phase in the case of intermediate
coupling. The low energy quasi-particle spectrum in g2SC phase contains four
gapless and only two gapped modes, and this phase has rather unusual 
properties at zero as well as at finite temperature. 

Under global charge neutrality condition, assuming that 
the effect of Coulomb forces and the surface tension is small, 
one can construct a mixed phase composed of positive charged 2SC phase 
and negative charged normal quark matter.

\vspace{1cm}

{\bf Acknowledgements}

M. H would like to thank the organizers of the workshop for invitation and 
offering a financial support. The work of M. H. was supported by the 
Alexander von Humboldt-Foundation, and the NSFC under Grant Nos. 10105005, 
10135030. The work of I.A.S. was supported by Gesellschaft 
f\"ur Schwerionenforschung (GSI) and by Bundesministerium f\"ur 
Bildung und Forschung (BMBF).

\end{document}